\documentclass[10pt,journal,cspaper,compsoc]{IEEEtran}
\ifCLASSOPTIONcompsoc
\else
\fi
\ifCLASSINFOpdf
   \usepackage[pdftex]{graphicx}
\else
   \usepackage[dvips]{graphicx}
\fi
\usepackage{algorithm}
\usepackage{algorithmic}
\ifCLASSOPTIONcompsoc
\usepackage[tight,normalsize,sf,SF]{subfigure}
\else
\usepackage[tight,footnotesize]{subfigure}
\fi
\begin{document}
%
\title{Constructing Limited Scale-Free Topologies Over Peer-to-Peer Networks}
%
%
%
%

\author{Eyuphan~Bulut,~\IEEEmembership{Member,~IEEE,}
        and~Boleslaw~K.~Szymanski,~\IEEEmembership{Fellow,~IEEE}
\IEEEcompsocitemizethanks{\IEEEcompsocthanksitem Eyuphan Bulut is with Cisco Systems, 2200 President George Bush Highway, 
Richardson, TX 75082, USA. He was with the Department
of Computer Science, Rensselaer Polytechnic Institute, Troy, NY, 12180, when he started this work. \protect\\
E-mail: ebulut@cisco.com
\IEEEcompsocthanksitem Boleslaw K. Szymanski (szymansk@cs.rpi.edu) is with the Department
of Computer Science, Rensselaer Polytechnic Institute, Troy, NY, 12180.
\IEEEcompsocthanksitem The very preliminary version of this paper is published in~\cite{netscicom}.}
\thanks{}}

%
%

\markboth{Transactions on Parallel and Distributed Systems,~Vol.~X, No.~X}%
{Bulut \MakeLowercase{\textit{et al.}}: Efficient Limited Scale-Free Overlay Topologies for Unstructured Peer-to-Peer Networks}
%


\IEEEcompsoctitleabstractindextext{%
\begin{abstract}
Overlay network topology together with peer/data organization and search 
algorithm are the crucial components of unstructured peer-to-peer (P2P) networks as they directly affect the efficiency of search on such networks. Scale-free (power-law) overlay network
topologies are among structures that offer high performance for these networks. A key problem for these topologies is the existence of hubs, nodes with high connectivity. Yet, the peers in a typical unstructured P2P network may not be willing or able to cope with such high connectivity and its associated load. Therefore, some hard cutoffs are often imposed on the number of edges that each peer can have, restricting feasible overlays to limited or truncated scale-free networks. In this paper,
we analyze the growth of such limited scale-free networks and propose two different algorithms for constructing perfect scale-free overlay 
network topologies at each instance of such growth. Our algorithms allow the user to define the desired scale-free exponent ($\gamma$). They also induce
low communication overhead when network grows from one size to another. Using extensive simulations, we demonstrate that these algorithms indeed generate perfect scale free networks (at each step of network growth) that provide better search efficiency in various search algorithms than the networks generated by the existing solutions.
\end{abstract}

}

\maketitle

\IEEEdisplaynotcompsoctitleabstractindextext

%
\IEEEpeerreviewmaketitle

\section{Introduction}
\label{sec:intro}

One of the significant properties of decentralized P2P networks is the topological characteristics of the formed overlay network topology. 
In addition to the distribution of data to peers and the type of search algorithm used, performance of search queries issued by peers is profoundly affected by overlay topology (i.e. logical connectivity graph). 
It has been shown~\cite{hui} that the performance is among the highest when the overlay topology is scale-free or has power-law degree distribution. This is because the network diameter of such topologies is small as it scales logarithmically (from $O(\ln N)$ to $O(\ln \ln N)$) with network size~\cite{ultra-small}.

Even though small-world or scale-free topologies offer efficient search, 
preserving these properties in growing distributed P2P environments is challenging. 
There have been many efforts to build such scale-free overlay structures for P2P networks in a centralized manner. However such centralized solutions are not scalable due to the difficulty of obtaining and maintaining global knowledge in a central node. 
Therefore, recently some algorithms using only the locally available information (i.e. neighbor peer's information) have been proposed. 
However, as expected, this has caused the loss of scale-freeness in the generated overlay structures, resulting in degradation of search efficiency.

A key difficulty in implementing scale-free networks is the existence of hubs. Nodes may not be willing or able to host such hubs because of excessive bandwidth and processing requirements resulting from high connectivity.  
Therefore, to promote fairness and topology acceptability~\cite{guclu}, some hard cutoffs are often imposed on the degree of each peer, making the topology a {\it limited scale-free network}. Clearly, these hard cutoffs might limit the scale-freeness of the entire topology. When the hard cutoff limit is lowered, the diameter of the network increases, reducing the search efficiency.

The construction of scale-free topologies with hard cutoffs and the effects of hard cutoffs on the search efficiency was first studied by Guclu et al. in~\cite{guclu}. However, the algorithm HAPA proposed by them has some deficiencies.
It is only a limited version of the well-known Barabasi-Albert (BA)~\cite{ba} or preferential attachment algorithm and with fully localized information, it
does not produce a perfect scale-free distribution of node degrees. Moreover, its time to converge and messaging overhead grow as the number of peers in the network increase. Even though it does not require global topology information at the time when nodes join, it needs total node count in the network available at each node, 
incurring some maintenance cost. Furthermore, it does not allow the user to set the desired scale-free exponent ($\gamma$), which significantly affects performance of search algorithms. 

In this paper, we address the challenges of growing scale-free overlay topologies with hard cutoffs. We first analyze the growth of limited scale-free networks. Then, based on this analysis, we propose two different algorithms that construct scale-free overlay network topologies having the following properties:
\begin{itemize}
\item{\it High adherence to scale-freeness:} Since the performance of the applications depends on the characteristics (i.e. diameter) of the overlay network on which they are built, the closer these networks adhere to the scale-free property, the more benefit the applications get from scale-free features.
\item{\it Parametrized:} Desired post-construction parameters (i.e. $\gamma$ that defines the search performance) of the scale-free network can be defined by the user.
\item{\it Practical:} Proposed models work on the growth of limited scale-free networks even if there is a limit on the number edges that a node can have in practice.
\item{\it Cost-Efficient:} An effective communication between newly joining nodes and existing ones during the construction of overlay topology with low communication overhead decreases traffic intensity and increases construction efficiency.
\end{itemize}

The rest of the paper is organized as follows. In Section~\ref{sec:back}, we overview the related work. In Section~\ref{sec:analysis}, we present our analysis on the growth of limited scale-free networks, and in Section~\ref{sec:proposed} we continue with the details of the proposed growth models. 
Section~\ref{sec:simulation} presents the simulation results. Finally, we conclude and outline future work in Section~\ref{sec:conclusion}.

\section{Background}
\label{sec:back}

\subsection{Scale-free networks}

Since the discovery of the scale-free property, scale-free networks have attracted a great deal of research interest in many natural and artificial systems such as the Internet~\cite{internet} and scientific collaboration networks~\cite{collaborations}. In these networks, nodes are connected according to the power-law of node degree distribution. That is, the degree distribution of nodes does not depend on the number of nodes in the network. The probability that a node has degree $i$ is proportional to $P(i)\approx i^{-\gamma}$, where the exponent is often limited to the range $2\leq \gamma \leq 3$. However, in limited scale-free networks, only nodes with degrees smaller than the achievable maximum degree comply with this rule. We will elaborate on this later.

The growth models for scale-free networks have been extensively studied in the last decade in network science. The notion of `preferential attachment' is usually the core of these proposed models. It is equivalent to Yule process~\cite{yule}, which is used to model the distribution of sizes of biological taxa. Price~\cite{price} first applied this idea to growth of networks under a mechanism called `cumulative advantage'. The concept `preferential attachment' and its popularity as scale-free network models is after Barabasi and Albert's work~\cite{ba} which independently rediscovered the same growth model on the web.

In the Barabasi-Albert (BA) model~\cite{ba}, each joining node selects and connects to an existing node $j$ with a probability ($p(j)$ = $\left(d_j/{\sum_{i=1}^{n}d_i}\right)$) that is proportional to the existing node's current degree, $d_j$. Each joining node computes $p(j)$ for each existing node in the network and randomly selects $k$ of them to connect to. The network formed by the BA model produces a power-law degree distribution with $\gamma = 3$, thus $P(i) \approx i^{-3}$. There are also other models that compute $p(j)$ differently than BA model does. However, they all use `preferential attachment' rule. A complete review of such models is given in~\cite{doro}.


In practical applications of scale-free networks, there is often a hard cutoff on the degree of nodes. Therefore, in this paper we focus on the limited scale-free networks and study the growth models on such networks\footnote{Yet, by setting a cutoff threshold equal to the number of nodes in the network, our algorithms are able to grow scale-free networks without cutoff.}. This is different than most of the previous works which study the growth of scale-free networks with no hard cutoff limit.

Moreover, in previously proposed growth models, there is only one way of computing connection probability for each node, resulting in one, predefined $\gamma$ value. In this paper, we propose growth models which define the connection probability of new joining nodes to existing nodes according to the network parameters (e.g. $\gamma$) desired in the final network.

In the construction of a network topology, it is also important to do the construction efficiently. Even though the growth of scale-free topologies has been extensively studied, less focus has been given to the applicability and construction overhead of growth models. In a real network application (such as peer-to-peer networks), the growth of such scale-free overlay topologies may cause high communication overhead between nodes. Whenever a new node joins the network, it needs the current degree information of all nodes (global topology information) to compute $p(j)$ for each existing node $j$ to select nodes to which it will connect. 
Different than this working principle, {\it computing with time} ~\cite{computer-bks} proposes a mechanism in which nodes self-select themselves according to their fitness to the task at hand. Each node computes maximum time delay, $t_d$, proportional to the inverse of its fitness. It responds to the request in a time uniformly randomly selected from interval $(0,t_d)$. The benefit of this scheme is that the communication needed for selection is constant in the number of candidates ~\cite{computer-bks}. In~\cite{gaian-dynamic}, Bent et al. used this mechanism to select connections in a scale-free graph using the degree of nodes as fitness. When a new node joins the network, it sends a network-wide broadcast message (using flooding) to announce its presence. Once the new node starts receiving the responses from existing nodes, it connects to the first $k$ responders (since each new node connects only to $k$ of the existing nodes at its joining time). To reduce communication, each existing node that has already sent or forwarded $k$ different responses (of other nodes or its own) to the newly joining node can stop forwarding any other responses, since they will not have any chance to be selected for connection. 

\subsection{Scale-free P2P Overlay Networks}

One of the first algorithms offering scale-free overlay network topology for unstructured peer-to-peer networks is the LLR algorithm~\cite{llr}. It is a variant of the BA model where only the nodes in the vicinity of a new node try to connect to it. Even though this algorithm helps in decreasing the construction overhead and relaxes the necessity of whole topology information, it causes divergence from scale-free network topology. Other similar approaches include~\cite{self}, where a self organizing scale free topology is proposed, and~\cite{overlay}, where robustness and fragility of such networks have been studied.

The above studies either require the availability of global network information or cannot construct limited scale-free overlay networks. To the best of our knowledge, the study~\cite{guclu} by Guclu et al. is the first work that studies the construction of limited scale-free overlay topologies for unstructured peer-to-peer networks as well as the the impact of hard cutoff on the  efficiency of search algorithms. The authors propose algorithms for building limited scale-free overlay structures for peer-to-peer networks considering the locality in the preferential edge assignment. For example, in the Hop-and-Attempt preferential attachment (HAPA) algorithm, each new node joining the network first selects a random node and then attempts to connect to it. If it can not achieve connection (due to hard cutoff and preferential selection probability) or it needs more nodes to connect (to fill all its $k$ stubs), it selects a random neighbor of currently selected but ineligible node and again attempts to connect to it. This continues until the new node fills all its stubs. Even though this algorithm works locally, it still assumes that the nodes know the total node count ($n$) in the network. The new node selects a random number between 0 and 1 and connects to a visited node $j$ if that random number is less than $p(j) = d_j/\sum_{i=1}^{n}d_i$, where the denominator is indeed equal to $2nk$. Moreover, since $p(j)$s get smaller as $n$ increases, as we will show in the simulation section, this algorithm may cause a new node to send a connection attempt message to many nodes in the network before succeeding to fill all its stubs. As a result, it may sometimes incur cost higher than the cost of a network-wide broadcast message. In a similar work~\cite{adhoc}, Guclu et al. also studied the impact of adhocness (mobility of peers) on the network topology and search efficiency.


\section{Analysis}
\label{sec:analysis}

In this section, we analyze the growth of limited scale-free networks with $n$ nodes, each with the minimum degree $k$, the average degree of all nodes $2k$, and the  maximum degree (hard cutoff) $m$, with the degrees distributed according to the power law with exponent $\gamma$. By definition, the nodes with maximum degree form a separate group from other nodes in terms of degree distribution:
\begin{eqnarray*}
P(i) &=& ci^{-\gamma}~\mbox{for nodes with degree}~k\leq i<m\\
P(i) &=& 1-\sum_{i=k}^{m-1}{P(i)}~\mbox{for}~i=m
\end{eqnarray*}
where $c$ is a constant. Note that, in this definition, only the nodes that have not yet reached degree $m$ are guaranteed to comply with the power-law degree distribution. 
However, as we show later in this section, whenever $m\geq 3k$, there exists a unique value of $\gamma$ with which all node degrees have frequencies in agreement with the power law.

Our goal is to construct a topology that shows perfect adherence to scale-free property. Moreover, we want to achieve this without using any global information. We characterize such a graph by its parameters: $n$, $m$, $k$ and $\gamma$, defined above. It is easy to show that the following inequalities must hold: $m>2k$ (we excluded here the trivial case of $m=2k$ in which all nodes of the graph are of degree $m$, trivially satisfying the definition of power law distribution of node degrees), $\gamma > 0$, and $n>2k$. We are interested in generated graphs with the number of nodes in the range $2k<n\leq n_{max}$ and we assume that $n_{max}>>k$.

In this paper, we assume a constant integer $k$ for the number of edges added by each joining node. However, it is a matter of simple extension to have instead a vector [$k_i$] of expected frequencies with which $i$ edges are added with the newly added node, such that  $k = \sum_{i=1}^{m}i k_i$.

The three constants, $k$, $m$, $\gamma$ are independent of each other except that for certain values of $m$, and $k$, there is a lower bound for $\gamma$'s.

Let $n_i$ denote the number of nodes with degree $i$ in the network with $n$ nodes.
By enumeration of all nodes and edges:
\begin{equation}
\label{eq1}
n = \sum_{i=k}^{m}n_i~\mbox{and}~2kn = \sum_{i=k}^{m} i n_i
\end{equation}	

Substituting $n$ in the above equations, and taking $n_m$ out, we get:
\begin{equation}
\label{eq3}
n_m = \frac{1}{m-2k}\sum_{i=k}^{m-1} (2k-i) n_i
\end{equation}

The power law degree distribution yields:
\begin{equation}
\label{eq4}
n_i  = \frac{cn}{i^{\gamma}}~\mbox{for}~i<m
\end{equation}

Using Eq.~\ref{eq4} to substitute $n_i$ in Eq.~\ref{eq3}, we get:
\begin{equation}
\label{eq5}
n_m  = \frac{cn}{m-2k}\sum_{i=k}^{m-1}\frac{2k-i}{i^{\gamma}}
\end{equation}

Using enumeration of nodes ($n_m + \sum_{i=k}^{m-1}n_i = n$) with different node degrees, we can compute the constant $c$ as:
\begin{equation}
\label{eq6}
c = \frac{m-2k}{\sum_{i=k}^{m-1}\frac{m-i}{i^{\gamma}}}
\end{equation}

Note that in limited scale-free networks we cannot enforce the power-law distribution for the nodes with maximum degree $m$ because their frequency is defined by Eq.~\ref{eq5}. However, for a given $m$ and $k$, nodes with maximum degree will also have frequency defined by the power-law ($n_m=cn/m^{\gamma}$) if $\gamma$ satisfies:

\begin{equation}
\label{eq65}
\frac{m-2k}{m^{\gamma}}=\sum_{i=k}^{m-1}\frac{2k-i}{i^{\gamma}}
\end{equation}

Since $m>2k$, the left hand side of Eq.~\ref{eq65} is always positive, its derivative for $\gamma$ is $-\ln(m)(m-2k)/m^\gamma$ while its value approaches $(1-2k/m)m^{-\gamma+1}$ when $\gamma$ tends to infinity. The right hand side of this inequality can be initially negative, but for large $\gamma$ it must be positive. Its value approaches $k^{-\gamma+1}$ when $\gamma$ tends to infinity and it has the derivative $-\sum_{i=k}^{m-1}\ln(i)\frac{2k-i}{i^\gamma}$.
It is easy to show that the right hand side decreases slower than the left hand side and therefore at most one unique value of $\gamma$ can satisfy Eq.~\ref{eq65}. The unique solution exists if and only if for $\gamma$=0, the right hand side is smaller than the left hand side, $m-2k \geq 2k(m-k)-(m-1)m/2+k(k-1)/2$
which reduces to $(m-2k+1/2)^2 \geq k^2+k-1/2$ and since $k^2<k^2+k-1/4<(k+1/2)^2$ then we get $m\geq 3k$. Thus, only for $m$ greater or equal to $3k$, there exists a unique value of $\gamma$ for which the constructed graph will have power-law distribution of all node degrees (including the nodes with maximum degree $m$).

Now, we will work on the general case in which the frequency of nodes with maximum degree does not need to comply with the power-law distribution. To be independent of the graph size $n$, we will use frequency $f_i = n_i/n$ of nodes with degree $i$. Then, substituting $c$ in $n_i$ definition with Eq.~\ref{eq6}, we get:
\begin{equation}
\label{eq7}
f_i = \frac{m-2k}{i^{\gamma}\sum_{j=k}^{m-1}\frac{m-j}{j^{\gamma}}}~\mbox{for}~i<m
\end{equation}
and
\begin{equation}
\label{eq9}
f_m = 1- \sum_{i=k}^{m-1}f_i
\end{equation}

Eq.~\ref{eq7} and Eq.~\ref{eq9} express frequencies, $f_i$'s, as simple functions of $m$, $k$ and $\gamma$.

Let's consider now a growth of the graph from its size of $n$ nodes to the size of $n+1$ nodes. The added node has $k$ edges originating from it which are then connected to the existing nodes, so on average it increases by $1$ the number of nodes with degree $k$, i.e. $n'_k$=$n_k+1$.

Let $a_i$ denote the average number of nodes that increase their degree from $i$ to $i+1$ in one step of growth (so the number of nodes of degree $i$ decreases by $a_i$ while the number of nodes with degree $i+1$ increases by $a_i$) by connecting to a newly added node. Of course, each existing node can add at most one connection to a newly added node. Hence, we have:
\begin{equation}
f_k(n+1)=f_kn + 1 - a_k
\end{equation}
which yields:
\begin{equation}
\label{eq10}
a_k = 1- f_k
\end{equation}

Similarly,  $f_i$ = $a_{i-1} - a_i$ for $k<i<m-1$, so by induction:
\begin{equation}
\label{eq11}
a_i = 1- \sum_{j=k}^{i}f_j~\mbox{for}~k\leq i < m-1
\end{equation}

Finally:
\begin{equation}
\label{eq12}
a_{m-1} = f_m
\end{equation}

All frequencies must be positive. For that to hold\footnote{Extended details and proofs are presented in our technical report~\cite{RN-NEST-01-2011}.}, it is necessary and sufficient that $n_m$$\geq 0$ or $\sum_{i=k}^{m-1}\frac{2k-i}{i^{\gamma}}\geq 0$, which can be rewritten as:

\begin{equation}
\label{eq13}
2k\sum_{i=k}^{m-1}i^{-\gamma} \geq \sum_{i=k}^{m-1}i^{-\gamma+1}
\end{equation}

\begin{figure}
\centering
\includegraphics[width=6.0cm, height=3.6cm]{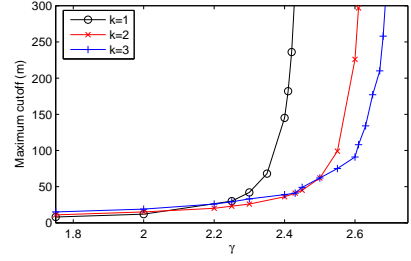}
\caption{Maximum cut off ($m$) values for different $k$'s in perfectly growing scale-free graph.}
\label{fig:maxm}
\end{figure}

If this condition is not satisfied, it is always sufficient either to appropriately increase $\gamma$ or $k$ or to sufficiently decrease $m$. Other changes to these parameters may or may not, depending on the particular values of the parameters, also cause the inequality of Eq.~\ref{eq13} to be satisfied. It is easy to notice that for $\gamma\geq 3$ this inequality is satisfied for arbitrary $m$ and $k$. Fig.~\ref{fig:maxm} plots the maximum values of $m$ for given $\gamma$ and $k$ values. It confirms that the maximum value of $m$ goes to infinity for $\gamma\geq 3$.

\begin{algorithm}
    \caption{AddNode(algoType, k, m)}
    \label{algAdd}
    \begin{algorithmic}[1]
    \STATE{n++}
    \IF{algoType = SRA}
    \STATE{numOfEdges = 0}
    \WHILE{(numOfEdges $<$ k)}
    \STATE{Pick a random number $r$ in [0,1)}
    \STATE{d[numOfEdges]= $\{$ d$|$ $v_{d-1}$ $\leq$ $r$ $<$ $v_{d}$ in Eq.~\ref{eqx}$\}$}
    \STATE{numOfEdges++}
    \ENDWHILE
    \STATE{Broadcast a message with $d$}
    \STATE{Add edge to the first $k$ responders}
    \ELSIF{algoType = SDA}
    \STATE{Broadcast a presence message}
    \STATE{All existing nodes receiving broadcast message increase total node count information by 1}
    \STATE{Nodes with degrees in SDA\_degList[n$\times$k] $\ldots$ SDA\_degList[n$\times$k+k-1] respond with their id and the total node count information}
    \STATE{Add edge to first responder from each degree}
    \STATE{Store the total node count information received from responders}
    \ENDIF
    \end{algorithmic}
\end{algorithm}

\section{Proposed Growth Models}
\label{sec:proposed}
We propose two different algorithms; (i) SRA and (ii) SDA. Algorithm~\ref{algAdd}, which is run at each node as the new nodes join the network, shows the steps of each algorithm. The first algorithm (SRA) needs one-time precomputation of v[] array using the formulations in previous section and the second algorithm needs $SDA\_degList[]$ array which is also computed one-time in Algorithm~\ref{alg1} before node joins start.

\subsection{Semi-Randomized Growth Algorithm (SRA)}
The rationale behind SRA (lines 2-10 in Algorithm~\ref{algAdd}) is to define the probability ranges (i.e., expected frequencies) for the selection of degrees of existing nodes to whom the new joining nodes will connect such that the final degree distribution of all nodes in the network will fit to the desired scale-free degree distribution. The algorithm starts with an initial configuration of a fully connected graph of $2k+1$ nodes. When a new node joins the network, it randomly decides the degrees of nodes that it will connect by generating $k$ random numbers, $r_1 \ldots r_k$, each in range [0,1] and finding the degree that each random number corresponds to. We define $v_i$'s (probability ranges) as:
\begin{eqnarray}
\label{eqx}
v_i &=& \left(v_{i-1}+\frac{a_i}{k}\right)~\mbox{$\forall$ $k \leq i \leq m-1$}
\end{eqnarray}
where $v_{k-1}=0$ and $v_i$'s are computed at each node one time only at the beginning (before node joins) using the $a_i$ formulas in the previous section. The random number $r_i$ corresponds to the degree $l$ such that $v_{l-1} \leq r_i < v_l$ is satisfied. Having computed all $k$ node degrees to which it wants to connect, the new node then broadcasts a message with these degree values. Once the existing nodes in the network receive such a message, the nodes of the desired degrees respond to the new node to establish a connection to it. Then, the new node selects the first $k$ of the nodes with desired degrees and connects to them. If the nodes with the desired degrees do not exist yet, which is likely only at the earlier stages of the network growth, until the first node reaches the degree $m$, the new node broadcasts a special request for the lower/higher degree nodes\footnote{Such a broadcast will be run only a limited number of times over initial small network, so its impact on communication overhead is negligible. Since initially nodes with degree $2k$ exist, the search will go in that direction.}, after the period of response for the original broadcast passes. Note that once $v$ array is known in advance, SRA will have the complexity of O(nmk) to grow a network of $n$ nodes. 

\subsection{Semi-Deterministic Growth Algorithm (SDA)}

\begin{algorithm}
    \caption{SDA\_ConnectionOrder(f[], k, m)}
    \label{alg1}
    \begin{algorithmic}[1]
    \FOR{(i=k-1;i$<$m;i++)}
    \STATE{freq[i] = f[i]/k}
    \STATE{n[i]=0}
    \STATE{score[i] = 2k $\times$ 2k $\times$ f[i]}
    \ENDFOR
  		\STATE{n[2k-1]=2k}
  		\FOR{(cur=k;cur$<$maxNodeCount;cur++)}
  		\FOR{(c=0;c$<$k;c++)}
   		\STATE{best = maxNodeCount}
  		\STATE{desired$\_$degree=k-1}
  		\FOR{(i=k-1;i$<$m-1;i++)}
  		\STATE{a1 = n[i-1]-score[i-1]-freq[i-1]}
  		\STATE{a2 = n[i]-score[i]-freq[i]}
  		\STATE{b1 = n[i-1]-1-score[i-1]-freq[i-1]}
  		\STATE{b2 = n[i]+1-score[i]-freq[i]}
  		\IF{i=k-1}
  		\STATE{a1++ a2++}
  		\ENDIF
  		\STATE{current = -$|$a1$|$-$|$a2$|$+$|$b1$|$+$|$b2$|$}
  		\IF{((best $>$ current) $\&$ (n[i]$>$0))}
  		\STATE{best = current}
  		\STATE{desired$\_$degree = i}
  		\ENDIF
  		\ENDFOR
  		\STATE{SDA\_degList[cur$\times$k+c] = desired$\_$degree+1}
  		\STATE{n[desired$\_$degree]$-$$-$}
  		\STATE{n[desired$\_$degree+1]++}
  		\FOR{(i = k-1; i$<$m;i++)}
  		\STATE{score[i] = score[i]+freq[i]}
  		\ENDFOR
  		\ENDFOR
  		\ENDFOR
  		\STATE{return SDA\_degList}
    \end{algorithmic}
\end{algorithm}

In the second algorithm, we aim to adhere as closely as possible to the desired scale-free network degree distribution at each step of the construction (i.e., for each intermediate network created during the construction). To this extend, $SDA\_degList[]$ array is precomputed one time using Algorithm~\ref{alg1} and used in the run time of SDA (lines 11-16 in Algorithm~\ref{algAdd}). In a network with $n$ nodes, for each edge of new joining node, Algorithm~\ref{alg1} decides which degree (of nodes) accepting such edge will minimize the divergence of the resulting degree distribution from the scale-free distribution. Let $n[i]$ denote current node count with degree $i+1$. $f[i]$ denotes the expected frequency of degree $i+1$ nodes (in perfect scale-free topology) as calculated in analysis section and $freq[i] = f[i]/k$ is the expected increment in degree-($i+1$) node count with only one edge addition to a new node. $score[i]$ denotes the current expected count of degree-($i+1$) nodes at given node count, and $score[i]+freq[i]$ denotes the expected count of degree-($i+1$) nodes after one new edge assignment to the new node. The algorithm starts with a fully connected $2k+1$ node graph and updates $score[i]$ values with each edge addition (line 29 in Algorithm~\ref{alg1}). With the $score[i]$, $freq[i]$ and $n[i]$ values known for the current network, the algorithm computes the sum of absolute differences between the current ($n[i]$) and expected ($score[i]+freq[i]$) counts of nodes with each degree that adheres to power law distribution. Let $f(i)=n[i]-score[i]-freq[i]$ show the difference of degree-$(i+1)$ nodes' expected count and current count in the network. Then, the sum of absolute differences for all degrees ($\sum_{k \leq j < m}{f(j)}$) in case of selecting degree $i+1$ can be computed as:
\[
d(i)=-|f(i-1)|-|f(i)|+ |f(i-1)-1|+ |f(i)+1|
\]  
Once the algorithm finds the degree that will decrease this sum the most (i.e., which provides the best adherence to scale-free distribution), it connects an edge from the new node to one of the nodes with such degree and updates the node counts (line 26-27 Algorithm~\ref{alg1}).

Note that the SDA algorithm deterministically finds the degree to which the new joining node should connect at each current total node count. However, the new node can select any node with such degree to connect (during run time), making the algorithm `semi-deterministic'. Once each node runs Algorithm~\ref{alg1} one time and gets `SDA\_degList[]' as output, they can decide their action (i.e., to respond or not for connection) every time they receive a new joining node's broadcast message in run time. They check the degree values in $SDA\_degList[n\times k]$ $\ldots$ $SDA\_degList[n\times k+k-1]$ using current node count $n$. If their degree is on the list, they respond to the new node with their ID and the total node count information in the network\footnote{In current setting, we assume fault-free communication between nodes.} such that the new node can also learn the current total node count in the network. The complexity of SDA is O(nmk)+O(nk), where the former is the cost of getting `SDA\_degList[]' from Algorithm~\ref{alg1} and the latter is the complexity of growing a network of $n$ nodes, making the overall SDA complexity O(nmk). Hence, the complexity of proposed algorithms matches the complexity of Gaian (O($nk$)) with constant $m$ and it is lower than the complexity of HAPA\footnote{The complexity of HAPA is not assessed in~\cite{guclu}, but the number of steps required in HAPA to find the next node to connect to grows with $n$, making the complexity higher than O($nk$).} and BA (O($n^2k$)).

\section{Simulation}
\label{sec:simulation}

In this section, we compare the proposed algorithms with well-known previous algorithms in terms of (i) the goodness of scale-free distribution, (ii) the effect of using global information vs. not using it, (iii) the search efficiency in different search algorithms, and  (iv) the messaging overhead and complexity incurred during the construction.  To this end, we study three different search algorithms: flooding (FL), normalized flooding (NF), and random walk (RW). In FL, source (i.e., query originating) node initiates the search by sending a message to its first hop neighbors. If the neighbor nodes receiving this message do not possess the requested item, they forward this message to their own neighbors, excluding the node from which they received the message. This type of forwarding process is repeated by each node that receives the message and does not have the requested item. In NF~\cite{nf}, a node receiving the query message only forwards it to $k$ (i.e., minimum degree of all nodes) of its neighbors in case it does not have the item in its repository. If a node has more than $k$ neighbors, it randomly selects only $k$ of them and forwards the message to them, excluding the one that sent the message to this node. Finally, in RW~\cite{rw}, a node receiving the query message and not possessing the requested item selects a single random neighbor and forwards the message to it. RW can also be considered as a special case of NF with a virtual minimum degree of 1. In all search algorithms, the forwarding of the message either stops after a predefined forwarding limit, which is called time-to-live (TTL), or the item is found at the current node. We also assume that search items are uniformly distributed among all nodes. Extensive details of these search algorithms can be found in~\cite{guclu}.


\begin{figure*}
\begin{center}
\subfigure[SDA $\gamma=2.5$]{\includegraphics[width=5.4cm, height=3.4cm]{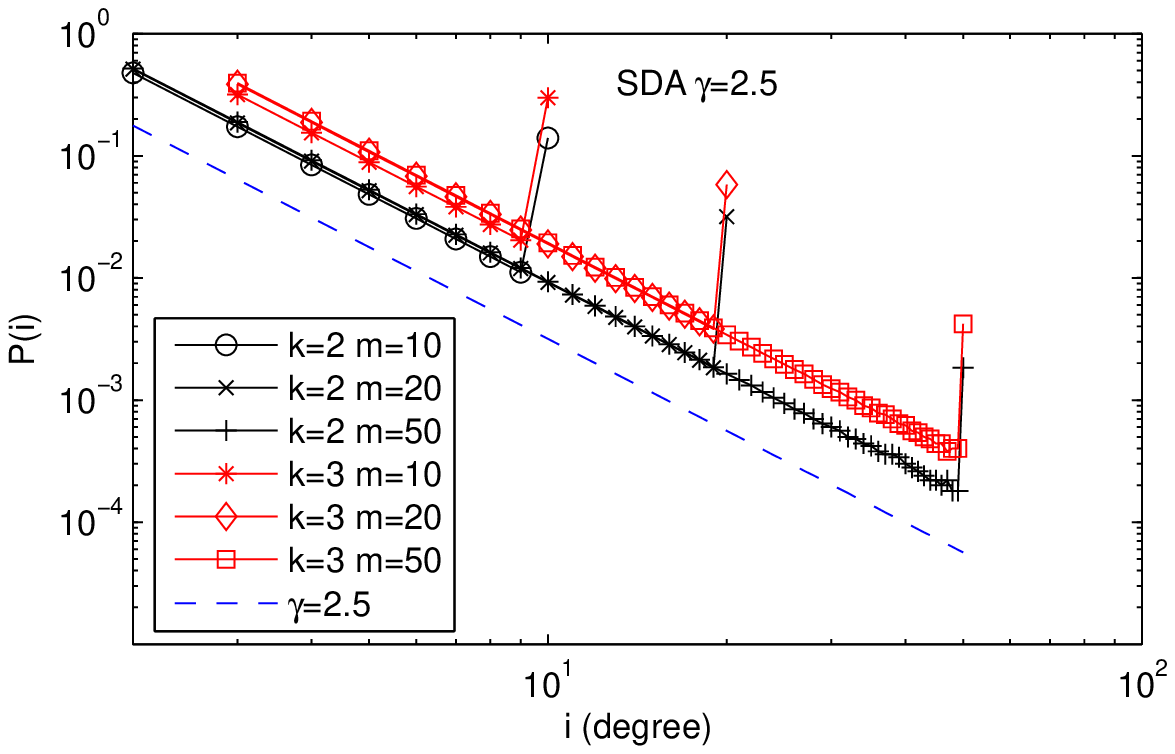}}
\subfigure[SDA $\gamma=2.7$]{\includegraphics[width=5.4cm, height=3.4cm]{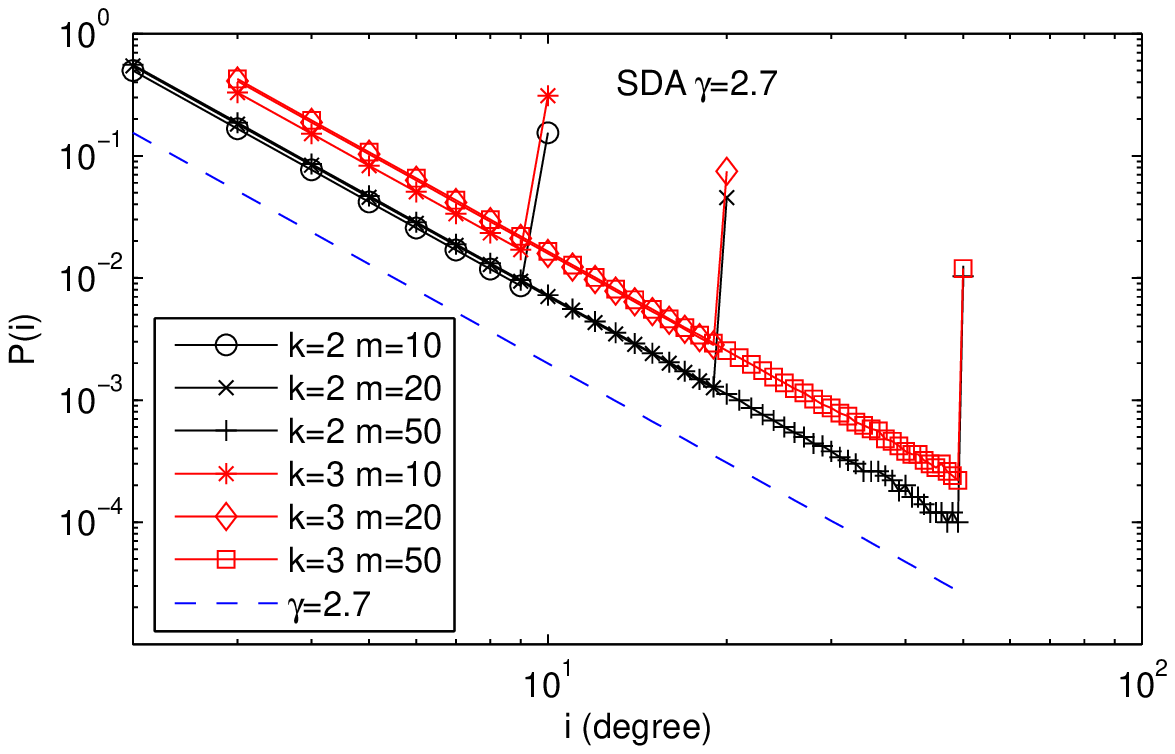}}
\subfigure[SDA $\gamma=3$]{\includegraphics[width=5.4cm, height=3.4cm]{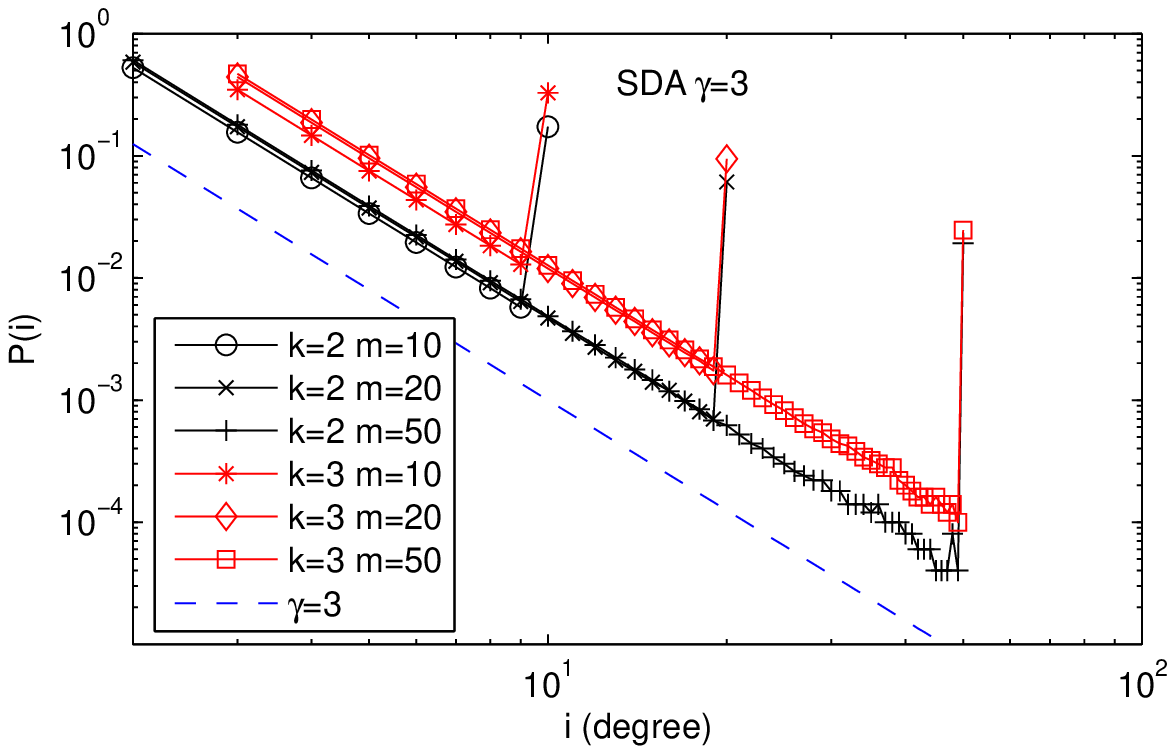}}\\
\subfigure[SRA $\gamma=2.5$]{\includegraphics[width=5.4cm, height=3.4cm]{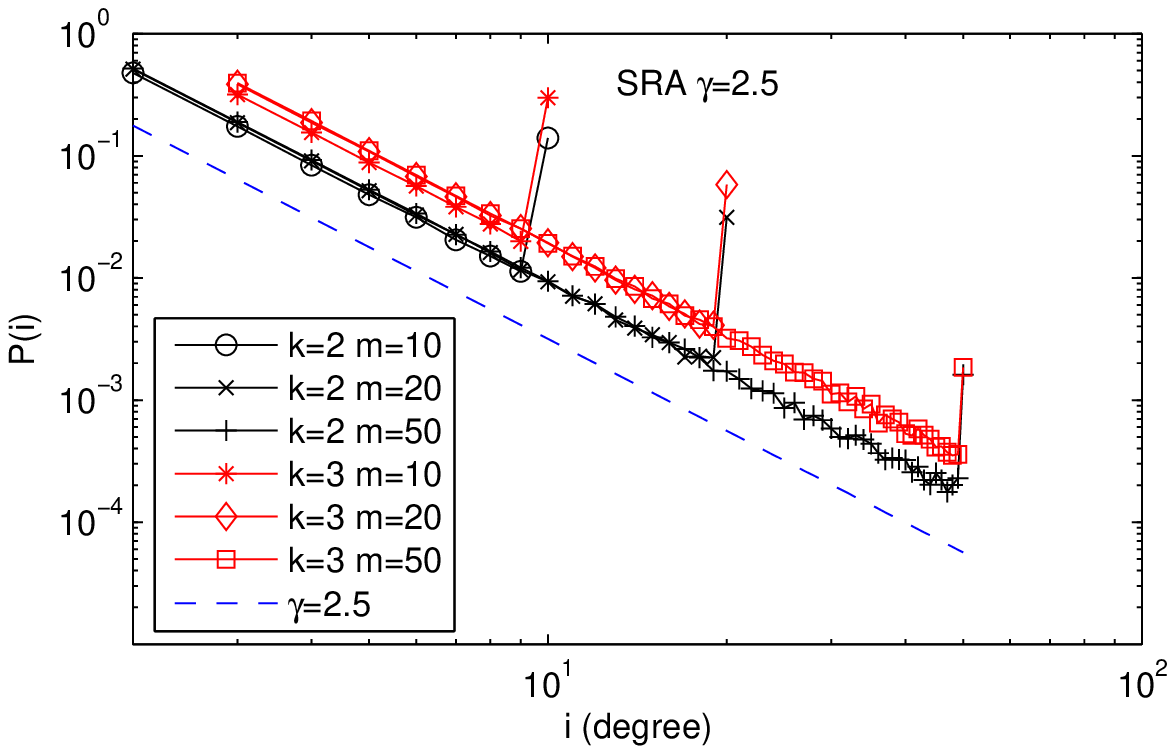}}
\subfigure[SRA $\gamma=2.7$]{\includegraphics[width=5.4cm, height=3.4cm]{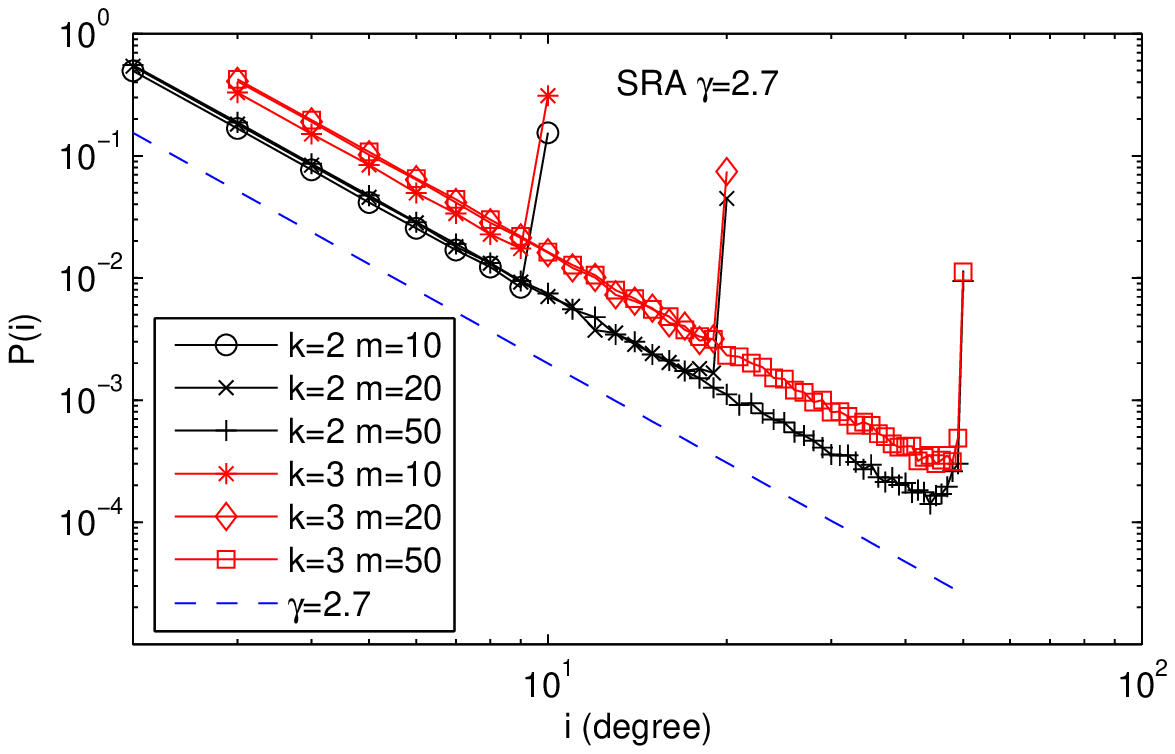}}
\subfigure[SRA $\gamma=3$]{\includegraphics[width=5.4cm, height=3.4cm]{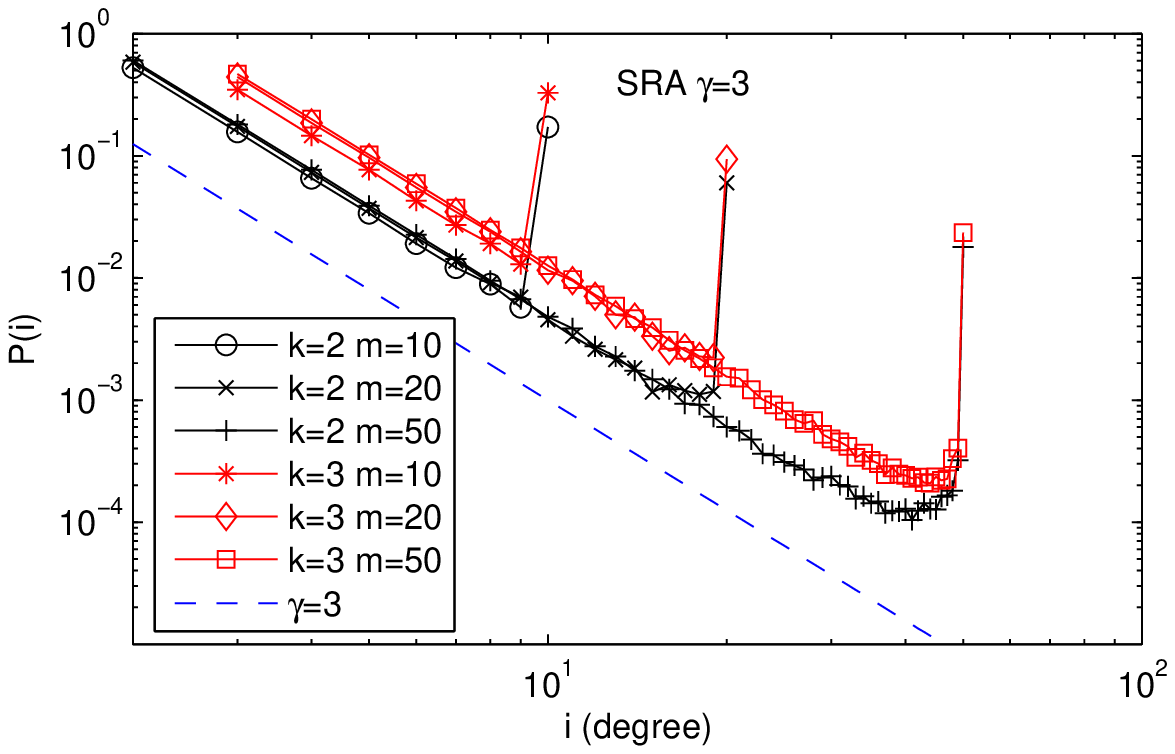}}\\
\subfigure[HAPA]{\includegraphics[width=5.4cm, height=3.4cm]{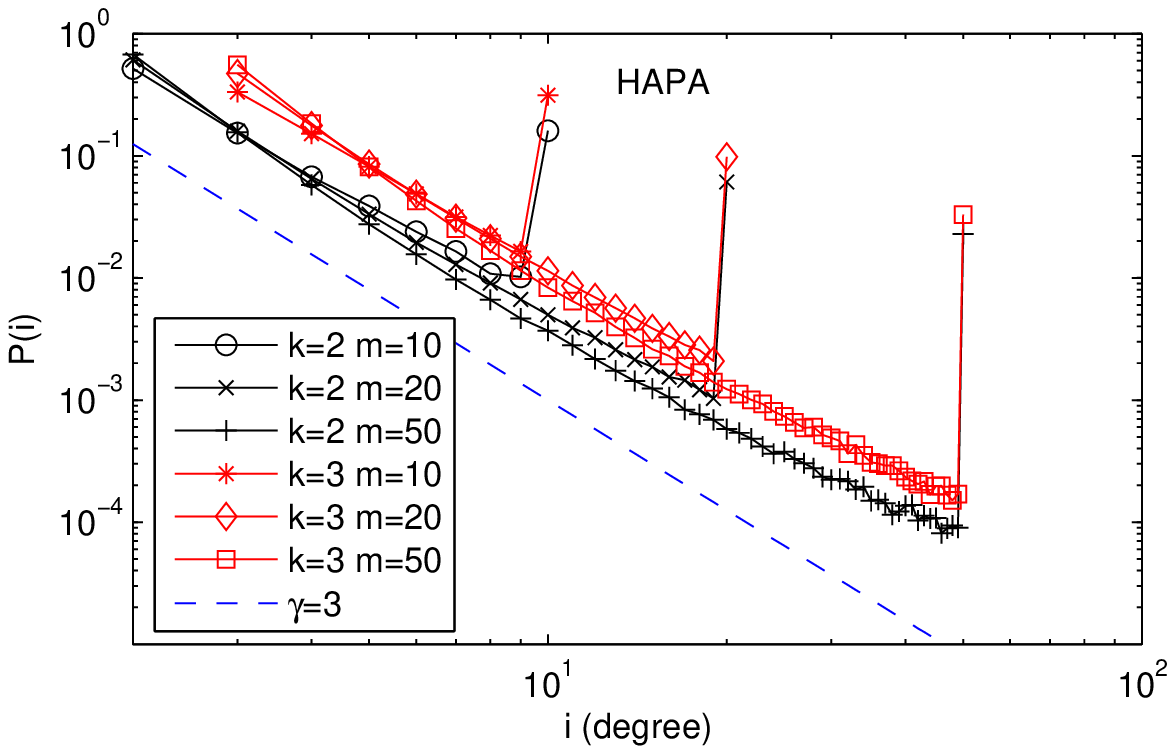}}
\subfigure[BA]{\includegraphics[width=5.4cm, height=3.4cm]{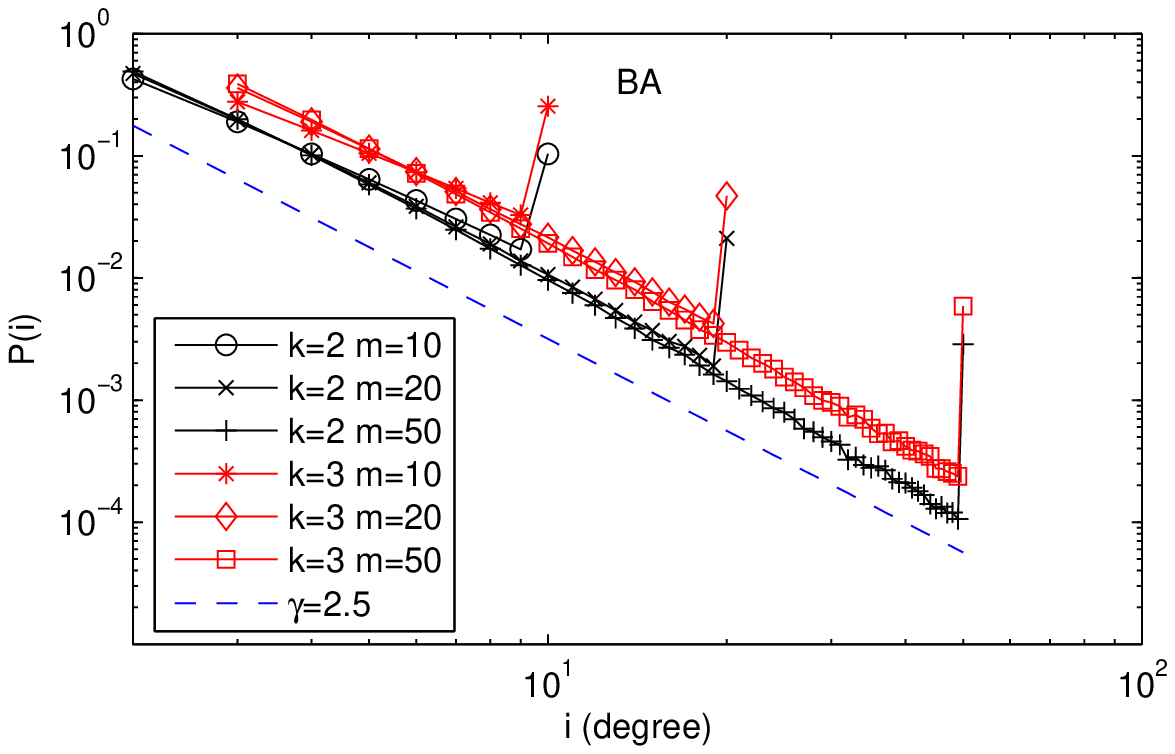}}
\subfigure[Gaian]{\includegraphics[width=5.4cm, height=3.4cm]{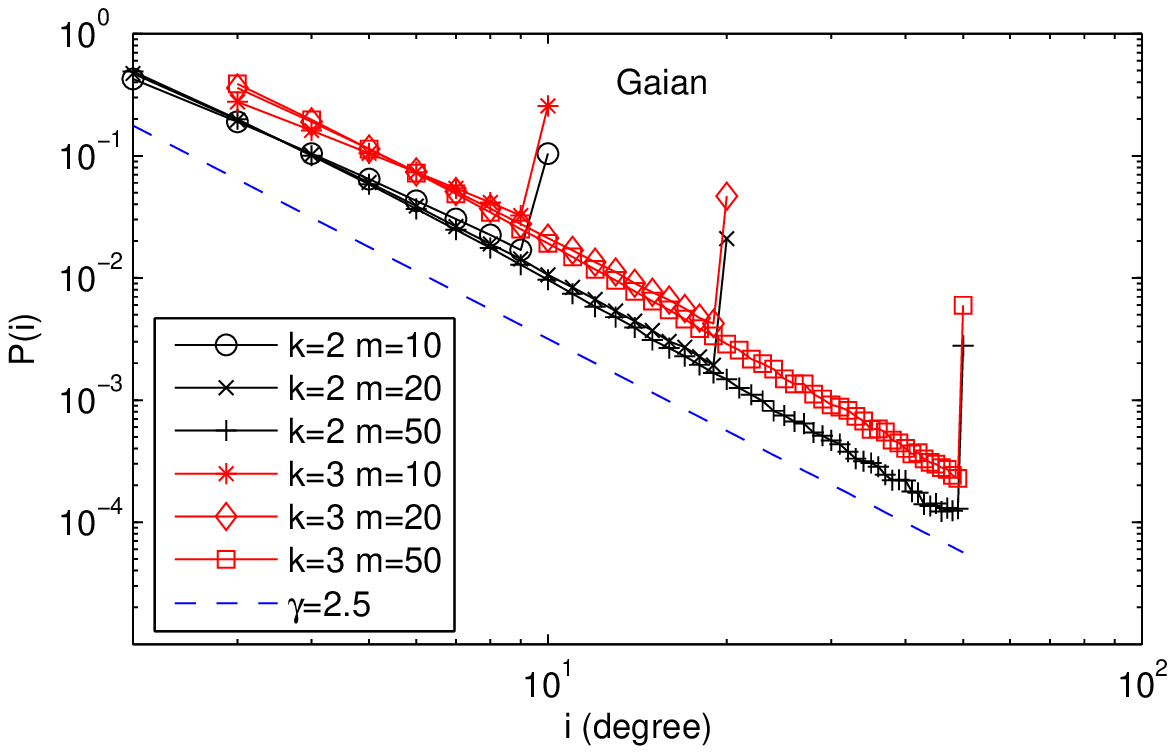}}
\caption{Degree distributions in different growth models ($n = 5$x$10^4$).}
\label{fig:dist}
\end{center}
\end{figure*}

\subsection{Simulation Results}

To compare the proposed growth model with existing algorithms, we generated different topologies (consisting of $n$ nodes) using different $k$, $m$ and $\gamma$ values. We start with a fully connected network of $2k+1$ nodes and add a new node to the network following the connection mechanisms of each growth model. The new node selects $k$ of the existing nodes (which have not reached its maximum edge limit) according to the algorithm in use and connects to them.

The algorithms we compare in the simulations are listed in Table~\ref{table1}. As the table shows, BA algorithm needs the global topology information (degrees of all nodes). Even though the HAPA algorithm does not need degrees of each node, it still needs the global knowledge of total node or edge count in the network. On the other hand, Gaian and our first algorithm, SRA, do not use any global knowledge. Our second algorithm, SDA, just uses the total node count, as does HAPA. While all other algorithms can only generate a network of fixed degree distribution exponent ($\gamma$) due to their designs, both of our algorithms can create topologies with desired exponent (so with desired network properties, such as diameter). 

\begin{table}[hbtp]
\begin{center}
\begin{tabular}{|l|p {3 cm}|p{2.5 cm}|}
\hline Algorithm & Global knowledge used & Flexible exponent ($\gamma$)\\ \hline \hline
BA~\cite{ba} &  Degrees of all nodes & No   \\ \hline
HAPA~\cite{guclu} &  Total node count & No \\ \hline
Gaian~\cite{gaian-dynamic} &  None & No    \\ \hline
SRA & None & Yes    \\ \hline
SDA & Total node count & Yes    \\ \hline
\end{tabular}
\end{center}
\caption{Comparison of Growth Models}
\label{table1}
\end{table}



\begin{figure*}
\centering
\subfigure[FL]{\includegraphics[width=5.4cm, height=3.4cm]{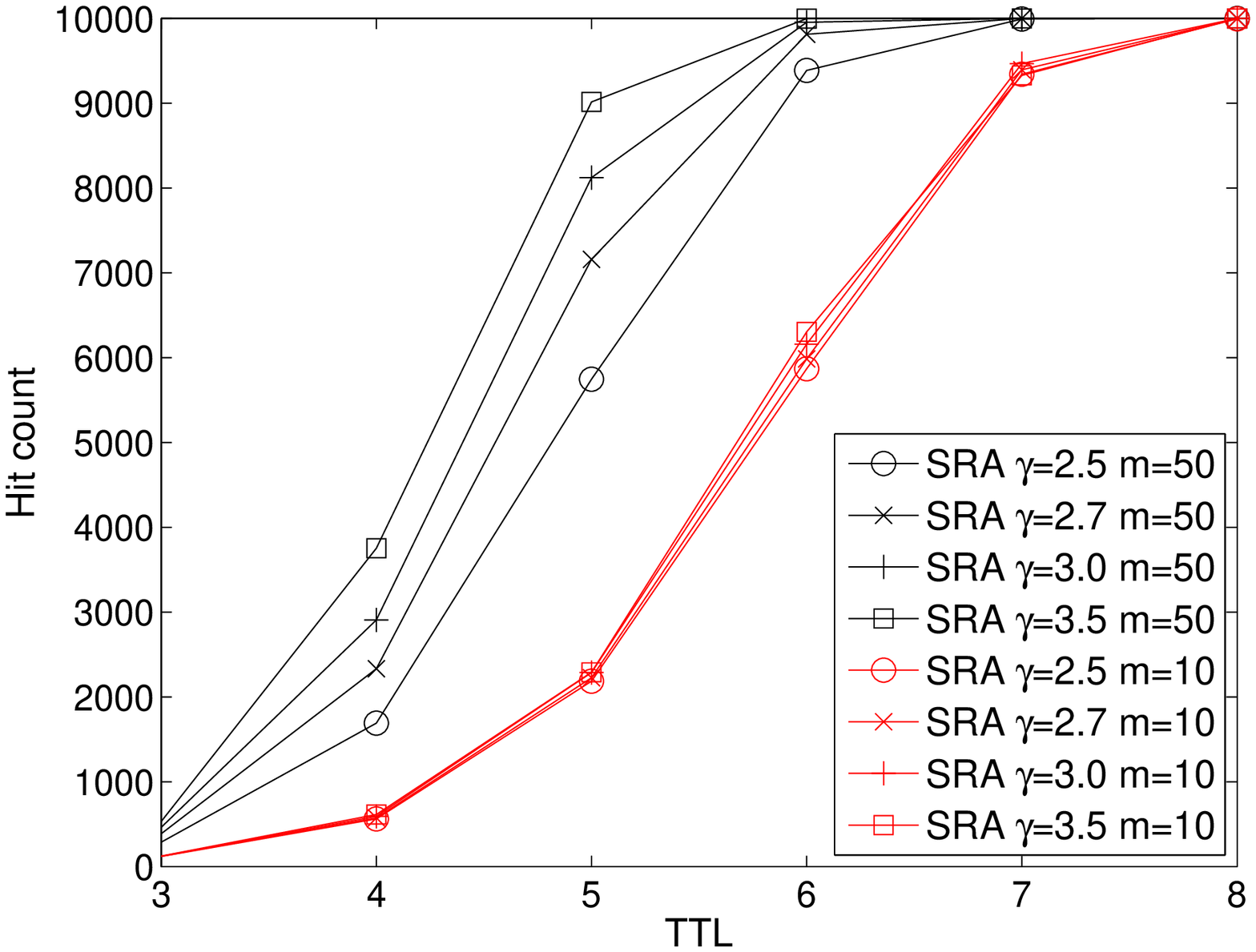}}
\subfigure[NF]{\includegraphics[width=5.4cm, height=3.4cm]{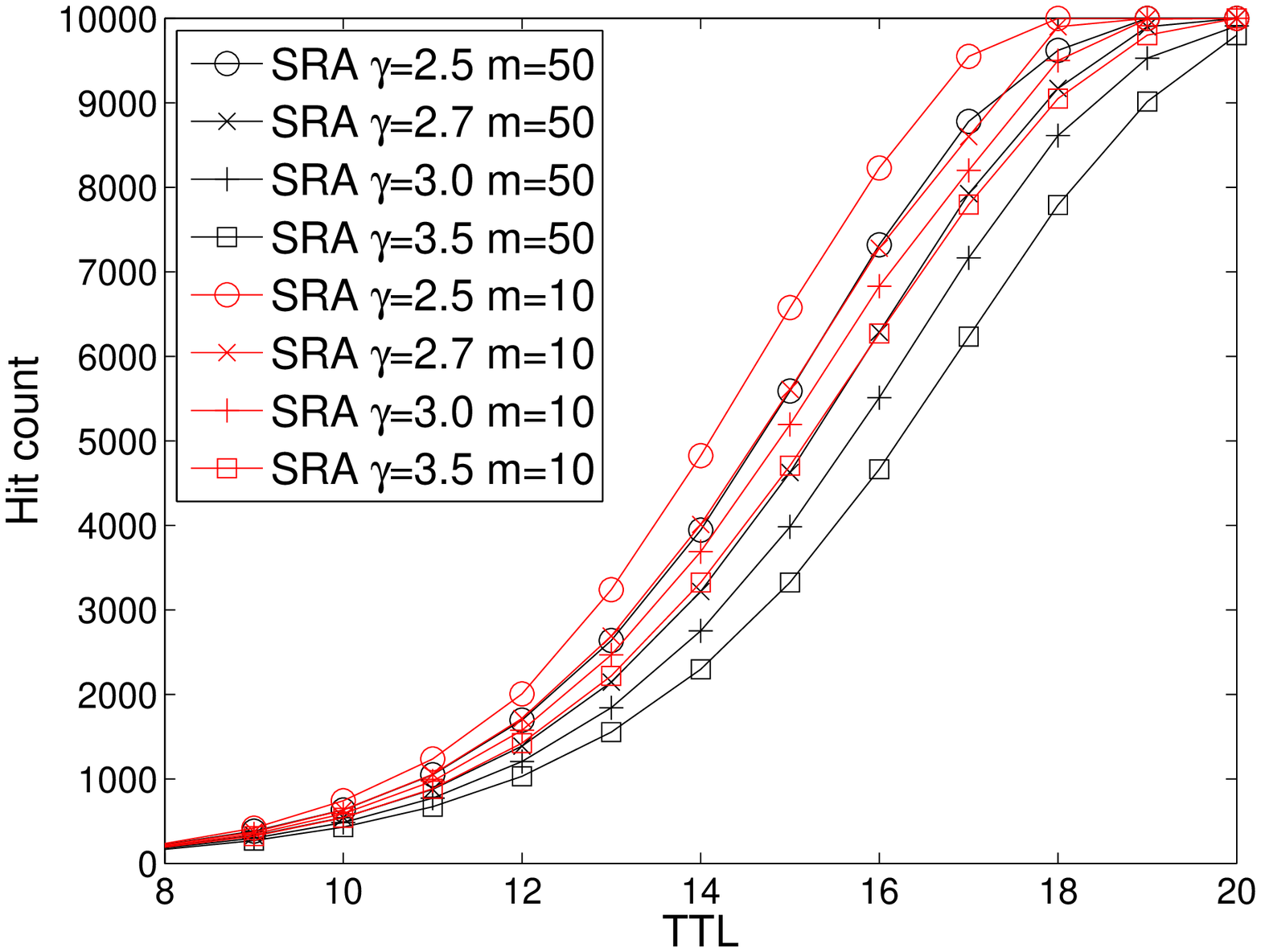}}
\subfigure[RW]{\includegraphics[width=5.4cm, height=3.4cm]{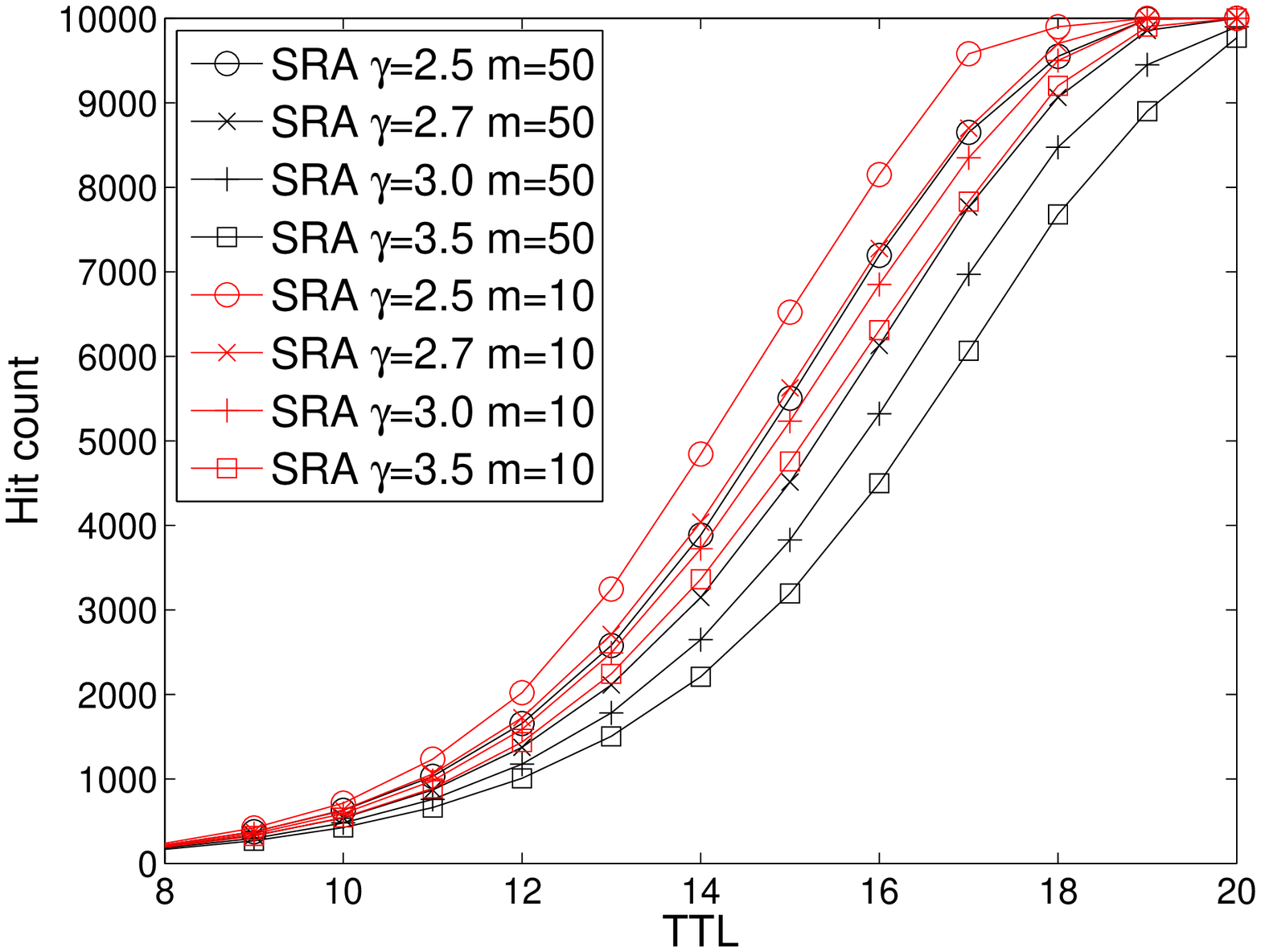}}
\caption{SRA search results}
\label{fig:sra}
\end{figure*}

In Fig.~\ref{fig:dist}, we show the degree distribution in topologies constructed by the compared algorithms. Since our algorithms can produce scale-free networks with a desired $\gamma$ exponent, we generated several network topologies with different $\gamma$ values. However, the other algorithms can yield a network only with a single $\gamma$ value. When we look at the degree distributions in topologies created by SDA in Fig.~\ref{fig:dist}a-~\ref{fig:dist}c, we clearly observe that the degree distributions perfectly match with the desired degree distribution of used $\gamma$ values in the construction. Similarly, there is a quite good match with the degree distributions of SRA algorithm and the predefined $\gamma$ value used in the construction. We only see a slight curve towards the end of the lines (high degree nodes) when $m=50$ in Fig.~\ref{fig:dist}e and Fig.~\ref{fig:dist}f. This is typically due to increasing impact of randomness used in the SRA algorithm and the insufficient number of nodes in the final network as the value of $\gamma$ increases. These minor curves disappear if the number of nodes in the network increases or the $\gamma$ value decreases (there is no such a curve in Fig.~\ref{fig:dist}d, where $\gamma$=2.5).

On the other hand, when we look at the degree distributions of the other algorithms, we can not see a good scale-free distribution even though some use global topology information during their construction phase. The degree distribution line is either concave or convex curve rather than a straight line. This result is indeed expected due to their designs which are originally proposed for general scale-free networks without hard cutoffs and later adjusted to limited scale-free networks. Thus, these figures clearly show that these algorithms are not able to create perfect limited scale-free topologies.

\begin{table}[hbtp]
\begin{center}
\begin{tabular}{|p{1.2cm}|p{0.9cm}|p{0.9cm}|c|c|c|}
\hline Method & SDA ($\gamma$=3.0)  & SRA ($\gamma$=3.0) & BA & HAPA & Gaian \\ \hline \hline
$\gamma$ ($m$=20) & 2.99187 & 2.97974 & 2.32319 & 3.04869 & 2.322818 \\ \hline
ChiSquare ($m$=20) & 0.00001 & 0.00064 & 0.00243 & 0.00435 & 0.00246 \\ \hline
$\gamma$ ($m$=50) &  3.00057 & 2.97269 & 2.46555 & 3.31755 & 2.465575 \\ \hline
ChiSquare ($m$=50) & 0.00007 & 0.00366 & 0.00494 & 0.01197 & 0.00498 \\ \hline
\end{tabular}
\end{center}
\caption{Results of fitness analysis}
\label{table:mle}
\end{table}

To quantify the fitting quality of distributions, we followed~\cite{powerlaw1} and used the maximum likelihood estimation (MLE) method to compute the best fitting $\gamma$ for results obtained from each algorithm. Then, using ChiSquare statistics, we computed the probability that the fit of data to power-law distribution with this $\gamma$ is random, so the smaller the value, the better the fit. Table~\ref{table:mle} shows the results for $k=2$ and $m\geq20$ \footnote{We used MLE for cases with $m\geq 20$ since it does not work well for small samples~\cite{powerlaw1}.}. The results show that SDA has superior performance, with the maximum divergence from the desired $\gamma$ of at most 0.008 and the very low probability, 0.00007, of the match being random. The next in performance is SRA with a divergence that is three times larger from $\gamma$ (0.028) and about 50 times higher in probability of a random match (0.00366). The compared methods have at least 8 (BA\footnote{Using the $\gamma = (3-2k/m)$ for limited BA derived from~\cite{guclu} gives even much (around 50 times) larger divergence.}, Gaian) to 16 (HAPA) times larger divergence, and probability of random fit from 70 to 170 times higher than SDA. The comparisons of performance with $k=3$, not shown for the sake of brevity, yielded similar results.


\begin{figure*}
\centering
\subfigure[FL $m$=10]{\includegraphics[width=5.4cm, height=3.4cm]{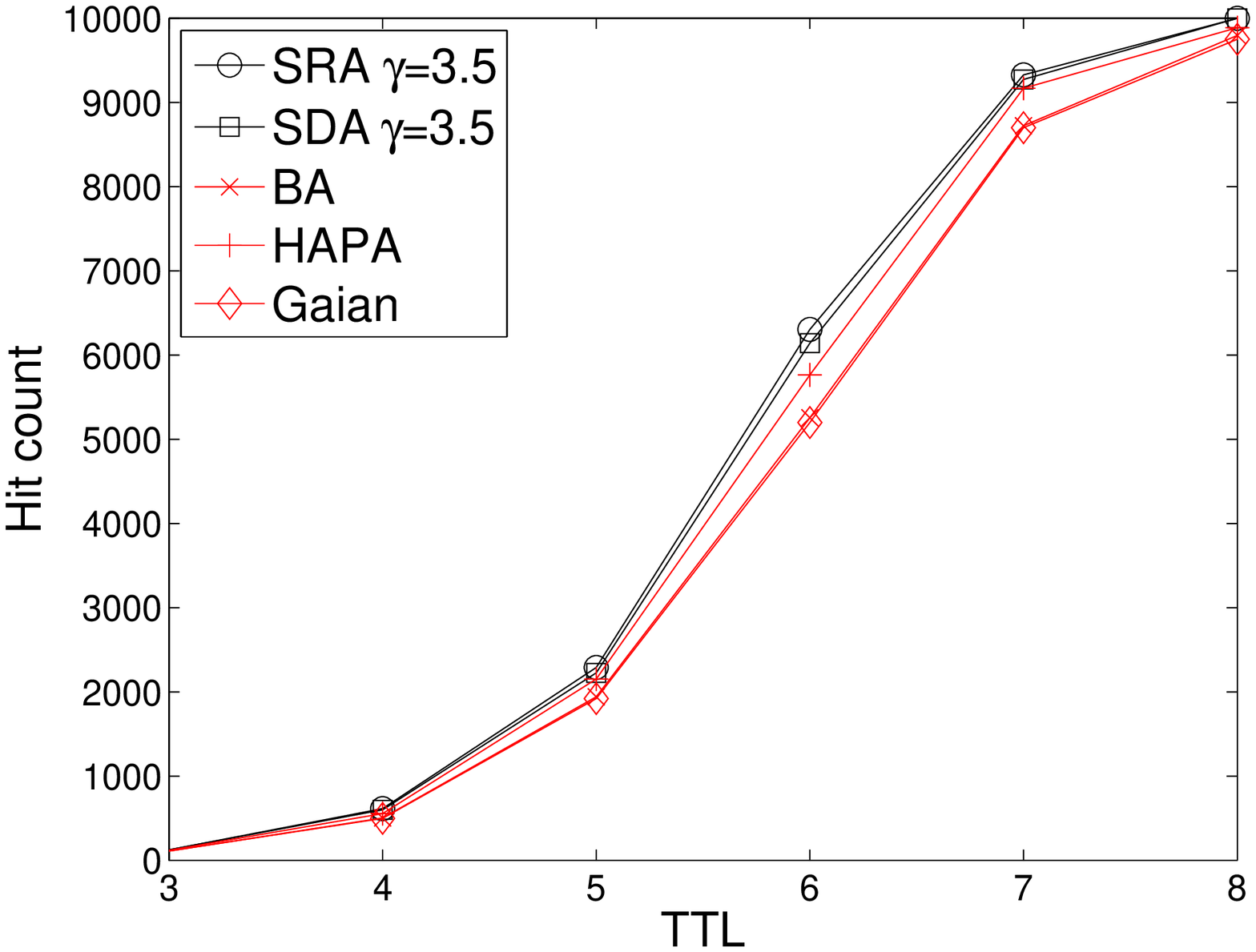}}
\subfigure[FL $m$=50]{\includegraphics[width=5.4cm, height=3.4cm]{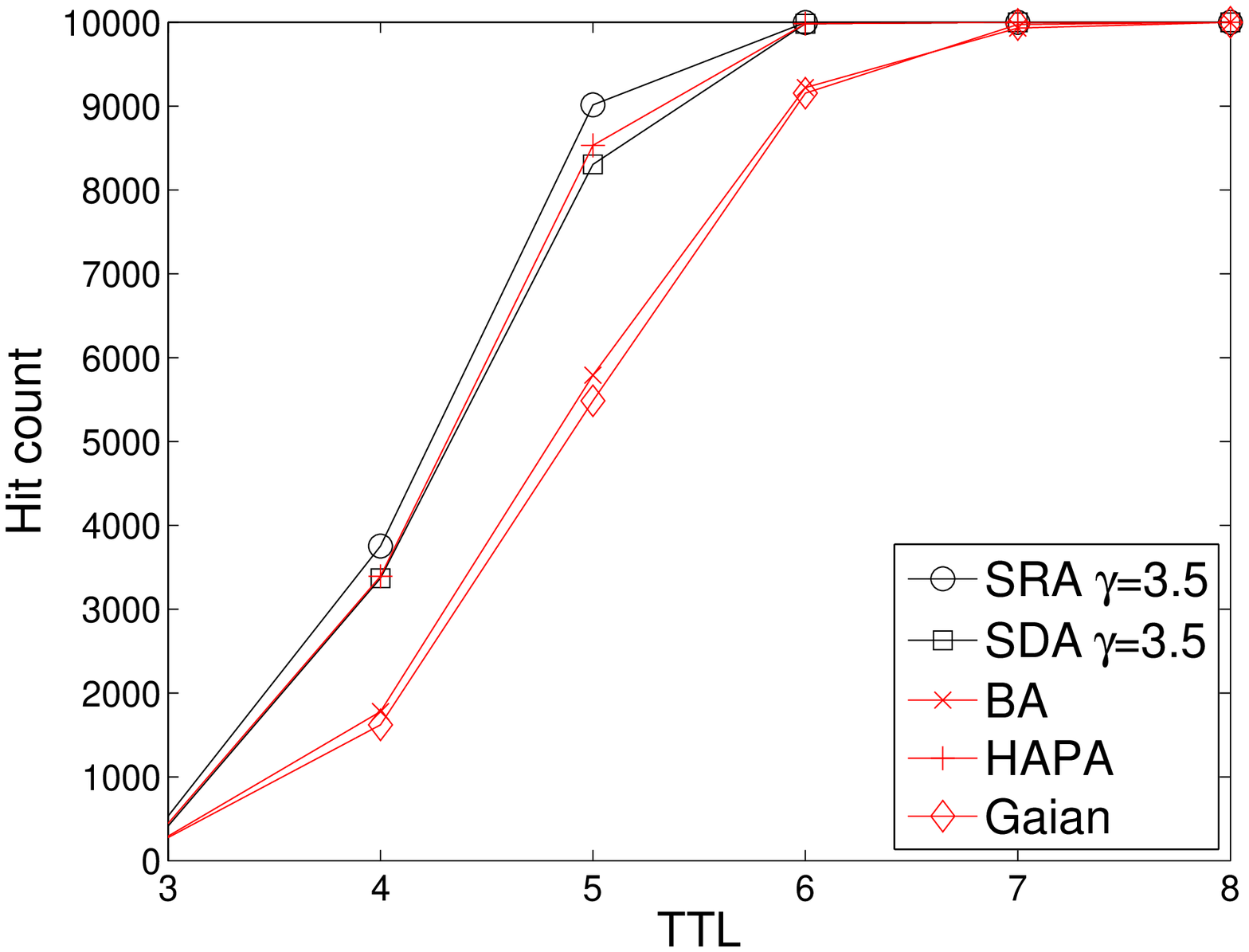}}
\subfigure[NF $m$=10]{\includegraphics[width=5.4cm, height=3.4cm]{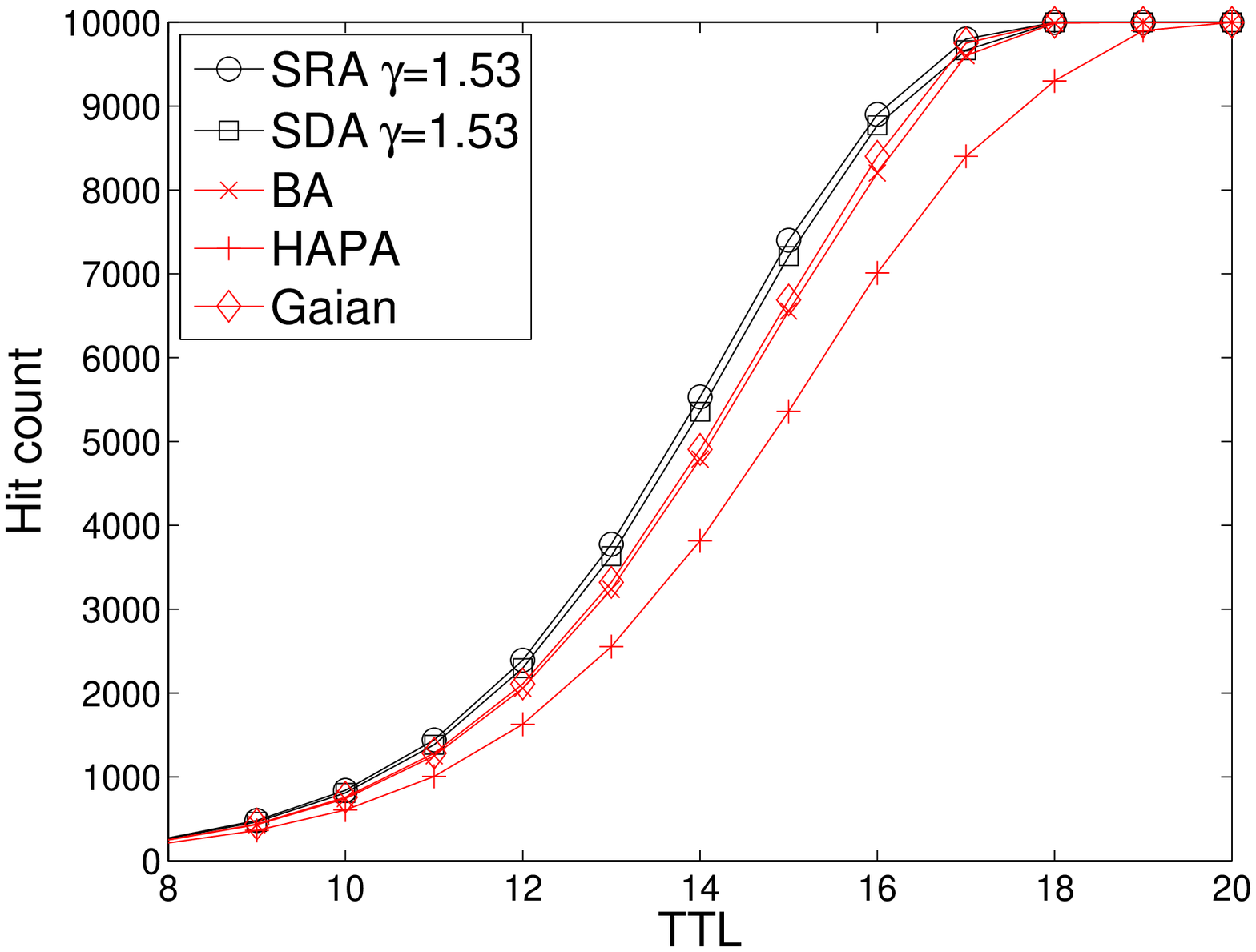}}\\
\subfigure[NF $m$=50]{\includegraphics[width=5.4cm, height=3.4cm]{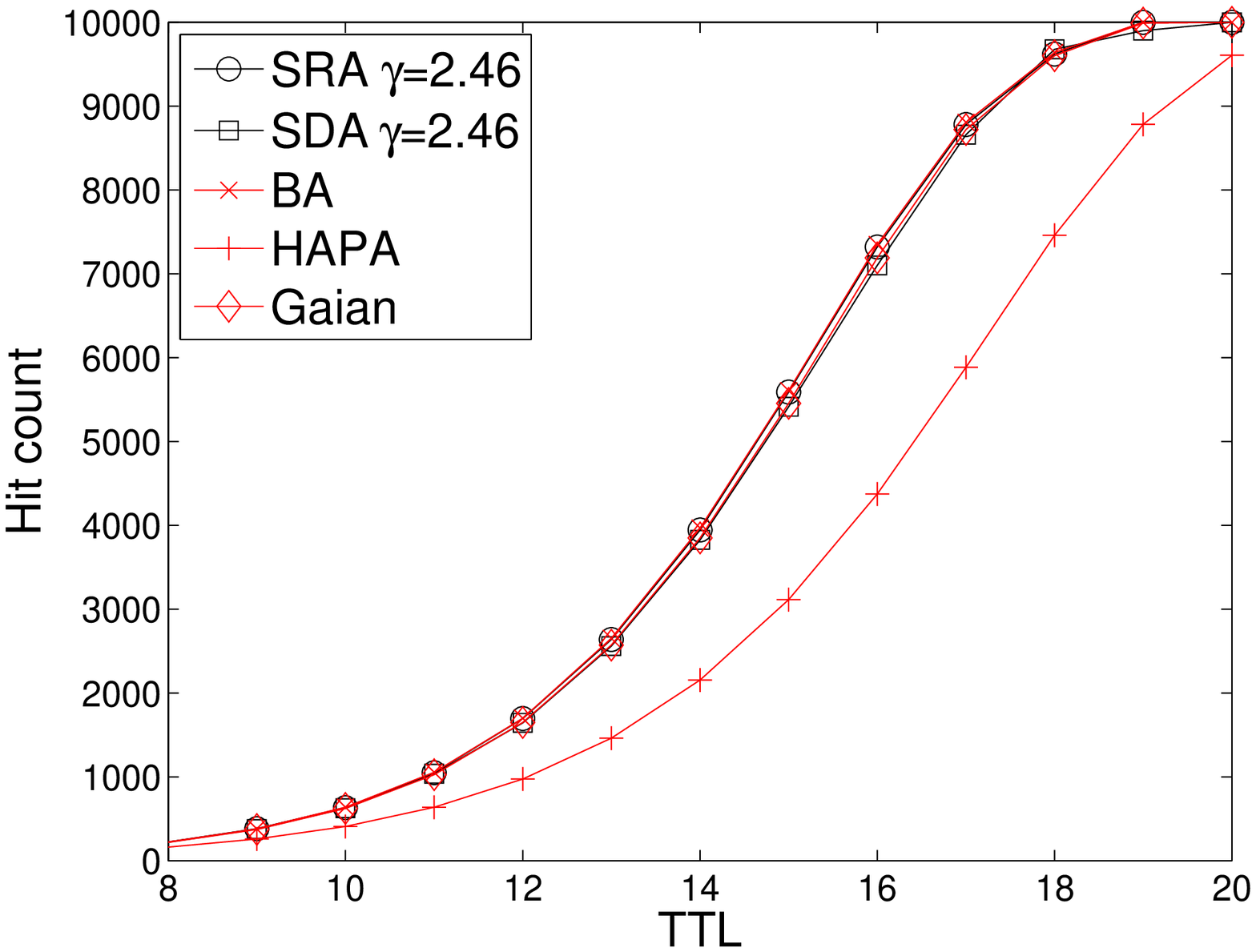}}
\subfigure[RW $m$=10]{\includegraphics[width=5.4cm, height=3.4cm]{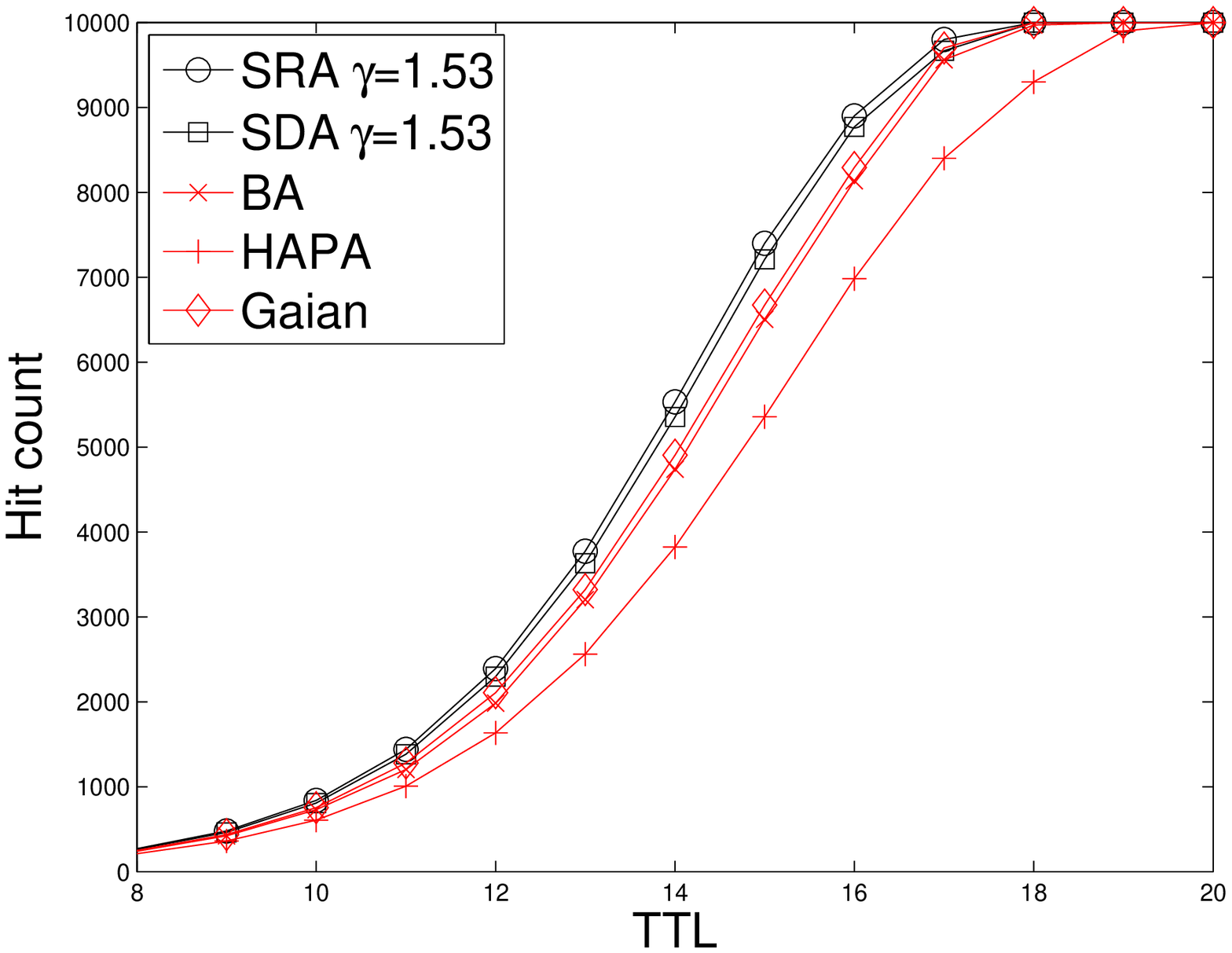}}
\subfigure[RW $m$=50]{\includegraphics[width=5.4cm, height=3.4cm]{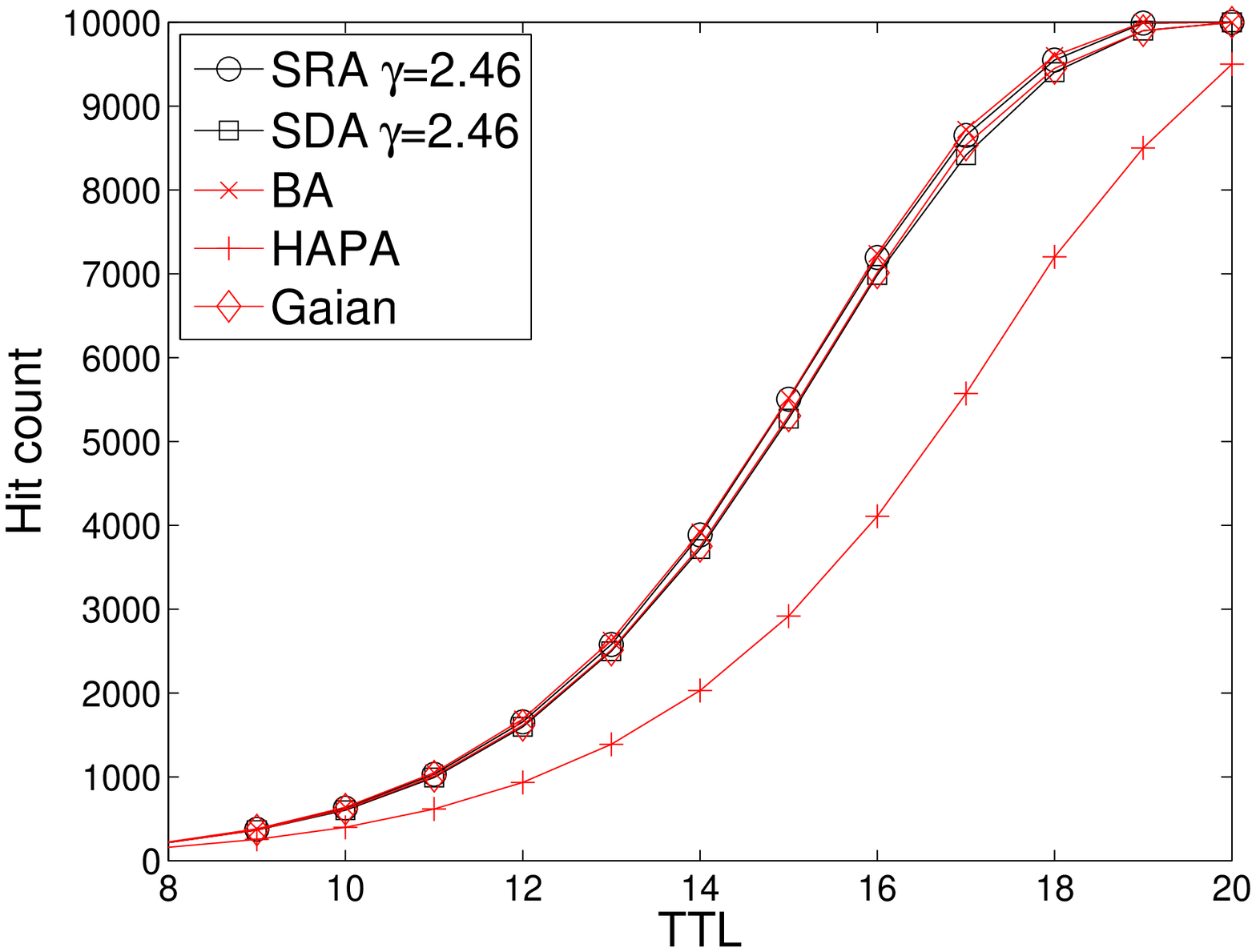}}
\caption{Search efficiency results: FL (a-b) NF (c-d) and RW (e-f)}
\label{fig:comparison}
\end{figure*}

Next, we look at the performance of proposed algorithms in terms of the search efficiency and compare their performance with other algorithms using different types of searches (in a network with $n=10,000$ nodes). As Fig.~\ref{fig:sra}a shows, the number of hits with given TTL value increases in SRA algorithm as the used $\gamma$ value increases. Moreover, the improvement becomes more visible as the limit on the degree of nodes, $m$, increases. On the other hand, Fig.~\ref{fig:sra}b-c show that as $m$ and $\gamma$ decreases, the NF and RW search efficiency increases, unlike the FL search performance. The impacts of $m$ and $\gamma$ on SDA algorithm is similar to the impacts of those parameters on SRA, thus, we did not present them for brevity. 

Fig.~\ref{fig:comparison}a-b show the comparison of FL search efficiency in all algorithms with $m$=10 and $m$=50, respectively. SRA algorithm with $\gamma =3.5$ achieves the best hit ratios with given TTL value in either case, while SDA with $\gamma$=3.5 (and HAPA algorithm when $m=50$) achieves the second best. BA and Gaian algorithms have much lower performance.


NF results in Fig.~\ref{fig:comparison}c-d reveal some interesting trends. Since our algorithms show the best NF search efficiency as $\gamma$ decreases, we set $\gamma=1.53$ (its lowest possible) when $m=10$ and $\gamma=2.46$ when $m=50$ (which also generates fully perfect scale-free topology where nodes with degree $m$ also comply the scale-free distribution). SRA/SDA algorithms have similar performance in both cases, and their performance is up to 12\% better than the performance of BA and Gaian algorithms when $m=10$ and similar to them when $m=50$. The reason why they cannot have better performance when $m=50$ is due to the larger minimum possible $\gamma$ (2.46) with given ($k=2$, $m=50$) setting. Note that HAPA has the worst NF search efficiency among all algorithms.


Fig.~\ref{fig:comparison}e-f show the RW based search results. Since RW needs more TTL to reach the destination node than NF, we applied the same normalization as used by previous work~\cite{guclu} to obtain the results of RW. That is, for a fair comparison we set the TTL value of RW search to the number of messages generated during NF search in the same setting. For example, the results in RW graphs with a TTL value of $x$ show the number of hits achieved by RW search with the same number of messages used in NF search that uses the same $x$ as the TTL. The comparison of algorithms in terms of RW based search efficiency leads to similar conclusions as the comparison of NF based search efficiency. We see that our algorithms perform better than all other algorithms when $m=10$ and similar to BA and Gaian algorithms and better than HAPA when $m=50$.

\begin{figure}
\centering
\includegraphics[width=5.7cm, height=3.5cm]{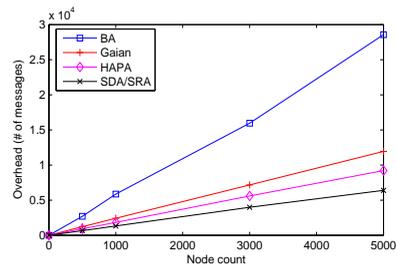}
\caption{Communication Overhead}
\label{fig:overhead}
\end{figure}

We also compared the algorithms in terms of the communication overhead (e.g. number of messages) during the construction of a scale-free network in Fig.~\ref{fig:overhead}. In all algorithms except HAPA, when a node wants to join the network it sends a broadcast message to announce its presence. Then, in BA algorithm every node sends its current degree count back to the new joining node. In Gaian algorithm, each node sends (or forwards) at most $k$ messages (containing degree of the corresponding node) towards the new joining node. This is also true in our algorithms, however, only nodes with desired degree respond, so the communication overhead is lower than it is in the Gaian algorithm. In HAPA algorithm, the new joining node first selects a random node and then attempts to connect to it. Next, it randomly walks in the network through neighbors until all its stubs are filled. Even though the HAPA algorithm is a localized algorithm, since each connection attempt by a new node becomes successful with probability $p(j)$ and only if the visited node has an edge count lower than the hard cutoff value, the new node's connection attempt message needs to travel a lot (sometimes a node is visited many times). Thus, it results in a large messaging overhead\footnote{The overhead of HAPA algorithm in Fig.~\ref{fig:overhead} does not include the overhead that will be generated for maintenance of total node count information at each node of the network. Its overhead will be higher if that would also be included. We could not include that cost since no detail is given about its implementation in~\cite{guclu}.}. Fig.~\ref{fig:overhead} clearly shows that the overhead of our algorithms is the smallest. Considering this result with the almost perfect degree distribution SRA achieves with a given exponent and without any global information, we can clearly state its superiority over other algorithms. The overhead of SDA is same to SRA and it achieves a much better fit to the scale-free property but it may not be robust during communication failures due to the requirement of total node count maintenance at every node.

\subsection{Summary of Contributions}

As the simulation results show, with the introduction of two new algorithms for P2P overlay networks, we demonstrated that:
\begin{itemize}
\item Without using any global information (i.e., SRA), we can construct a sequence of growing scale-free overlay topologies with lower overhead than the currently used algorithms. Each of these evolving topologies adheres nearly perfectly to the scale-free degree distribution. This is contrary to the statement in~\cite{guclu} indicating the necessity of some global information to achieve close scale-freeness. Moreover, the other algorithms can not show good scale-freeness, including those that use global information.
\item The effect of scale-free exponent, $\gamma$, is significant in achieving high search efficiency in different search algorithms (especially for weakly connected networks (i.e. $k$=2)). While our algorithms with $\gamma$=3.5 achieves the best hit ratios in FL search, they achieve the best hit ratios in NF and RW searches when $\gamma$ is set to lowest possible value ($\gamma=1.53$ ($m=10$) and $\gamma=2.46$ ($m=50$)). Since our algorithm can create a scale-free network with a desired $\gamma$ value, the value that gives the best search efficiency for the given search algorithm can be used to create the scale-free overlay topology to increase the performance. The other algorithms can only achieve high search efficiency either in FL (HAPA) or NF/RW (BA and Gaian).
\item In limited scale-free networks, when $m\geq 3k$, there is a unique $\gamma$ with which all node degrees in the sequence (including the ones with degree $m$) of networks of growing size comply with power law distribution.
\end{itemize}

\section{Conclusion}
\label{sec:conclusion}

In this paper, we introduced two new algorithms for growing limited scale-free overlay topologies for unstructured P2P networks. In extensive simulations we demonstrated clear superiority of our algorithms with well known algorithms in literature. Our algorithms provide almost perfect adherence to the scale-free property using zero or limited global information and require less communication during overlay construction than others do. They also provide higher search efficiency for different search methods when a network is constructed with right parameters (i.e., $\gamma$). In future work, we plan to develop algorithms which maintain the perfect scale-freeness without using global information while nodes join and leave the graph at the same time.

\ifCLASSOPTIONcompsoc
  \section*{Acknowledgments}
\else
  \section*{Acknowledgment}
\fi

Research was sponsored by US Army Research Laboratory and the UK  Ministry of Defence and was accomplished under Agreement Number W911NF-06-3-0001. Research was also sponsored by the Army Research Laboratory and was accomplished under Cooperative Agreement Number W911NF-09-2-0053. The views and conclusions contained in this document are those of the authors and should not be interpreted as representing the official policies, either expressed or implied, of the US Army Research Laboratory, the U.S.  Government, the UK Ministry of Defence, or the UK Government. The US and UK Governments are authorized to reproduce and distribute reprints for Government purposes notwithstanding any copyright notation hereon.

The authors also express their thanks to Ferenc Molnar from the NeST Center at RPI for computing the MLE and ChiSquare results for the test data.
\ifCLASSOPTIONcaptionsoff
  \newpage
\fi

\end{document}